\documentclass[a4paper]{jpconf}
\usepackage{graphicx}
\def\nuebar{\rm{\bar{\nu_e}}}

\begin{document}

\hfill AS-TEXONO/06-06\\
\hspace*{1cm} \hfill \today

\title{Research program towards observation of neutrino-nucleus
coherent scattering}

\author{H T Wong$^{1, \ast }$, H B Li$^1$, S K Lin$^1$, S T Lin$^1$,
D He$^2$, J Li$^2$, X Li$^2$, Q Yue$^2$, Z Y Zhou$^3$
and S K Kim$^4$
}

\address{
$^1$ Institute of Physics, Academia Sinica, Taipei 11529, Taiwan.\\
$^2$ Department of Engineering Physics, Tsing Hua University,
Beijing 100084, China.\\
$^3$ Department of Nuclear Physics,
Institute of Atomic Energy, Beijing 102413, China.\\
$^4$ Department of Physics,
Seoul National University, Seoul 151-742, Korea.
}

\ead{htwong@phys.sinica.edu.tw ($^\ast$Corresponding Author)}

\begin{abstract}
The article describes the research program 
towards an experiment to observe 
coherent scattering between neutrinos and the nucleus
at the power reactor.
The motivations of studying this process are surveyed. 
In particular, a threshold of 100-200~eV has been 
achieved with an ultra-low-energy germanium detector prototype.
This detection capability at low energy can also be adapted
for searches of Cold Dark Matter in the low-mass region
as well as to enhance the sensitivities in the study
of neutrino magnetic moments.
\end{abstract}

Neutrino coherent scattering with the nucleus\cite{coher}
\begin{equation}
\rm{
\nu ~ + ~ N ~ \rightarrow ~
\nu ~ + ~ N
}
\end{equation}
is a fundamental neutrino interaction
which has never been experimentally observed.
The Standard Model cross section for
this process is given by:
\begin{equation}
\label{eq::cohdsm}
\rm{
( \frac{ d \sigma }{ dT } ) ^{coh} _{SM}   = 
\frac{ G_F^2 }{ 4 \pi }
m_N  [ Z ( 1 - 4 sin^2 \theta_{W} ) - N ]^2
[ 1 -  \frac{m_N T_N }{2 E_{\nu}^2 } ] 
}
\end{equation}
\begin{equation}
\label{eq::cohsm}
\rm{
\sigma _{tot}  = 
\frac{ G_F^2  E_{\nu}^2 }{ 4 \pi }
 [ Z ( 1 - 4 sin^2 \theta_{W} ) - N ]^2  ~~,
}
\end{equation}
where $\rm{m_N}$, N and Z are the mass, neutron number
and atomic number of the nuclei, respectively,
$\rm{E_{\nu}}$ is the incident neutrino energy
and $\rm{T_N}$ is the measure-able
recoil energy of the nucleus.
This formula is applicable for $\rm{E_{\nu} < 50~MeV}$
where the momentum transfer ($\rm{Q^2}$) is small such that
$\rm{ Q^2 R^2 < 1 }$, where R is the nuclear size.
Although the cross-section is relatively large
due to the $\sim$N$^2$ enhancement by coherence,
the small kinetic energy from nuclear recoils
poses severe experimental challenges
both to the detector sensitivity and to background
control. 
Various detector techniques have been 
considered\cite{otherexpt} to meet these challenges.

Measurement of the coherent scattering
cross-section would provide a sensitive
test to the Standard Model\cite{smtest}, probing the
weak nuclear charge and radiative corrections
due to possible new physics above the weak scale.
The coherent interaction plays important role
in astrophysical processes where the
neutrino-electron scatterings are suppressed
due to Fermi gas degeneracy. It is significant
to the neutrino dynamics and energy transport
in supernovae and neutron stars\cite{astro}.
Being a new detection channel for neutrinos,
it may provide new approaches to study other aspects
of neutrino physics, such as that for 
supernova neutrinos\cite{nostos}.
Coherent scattering with the nuclei is also the
detection mechanism adopted in the
direct Dark Matter searches\cite{cdm},
such that its observations
and measurements with the
known particle neutrino is an
important milestone.
Furthermore, neutrino coherent scattering
may be a promising avenue towards a compact
and relatively transportable neutrino detector,
an application of which can be for
the real-time monitoring on the operation
of nuclear reactors\cite{monitor}.

\begin{figure}[hbt]
\begin{minipage}{18pc}
\includegraphics[width=16pc]{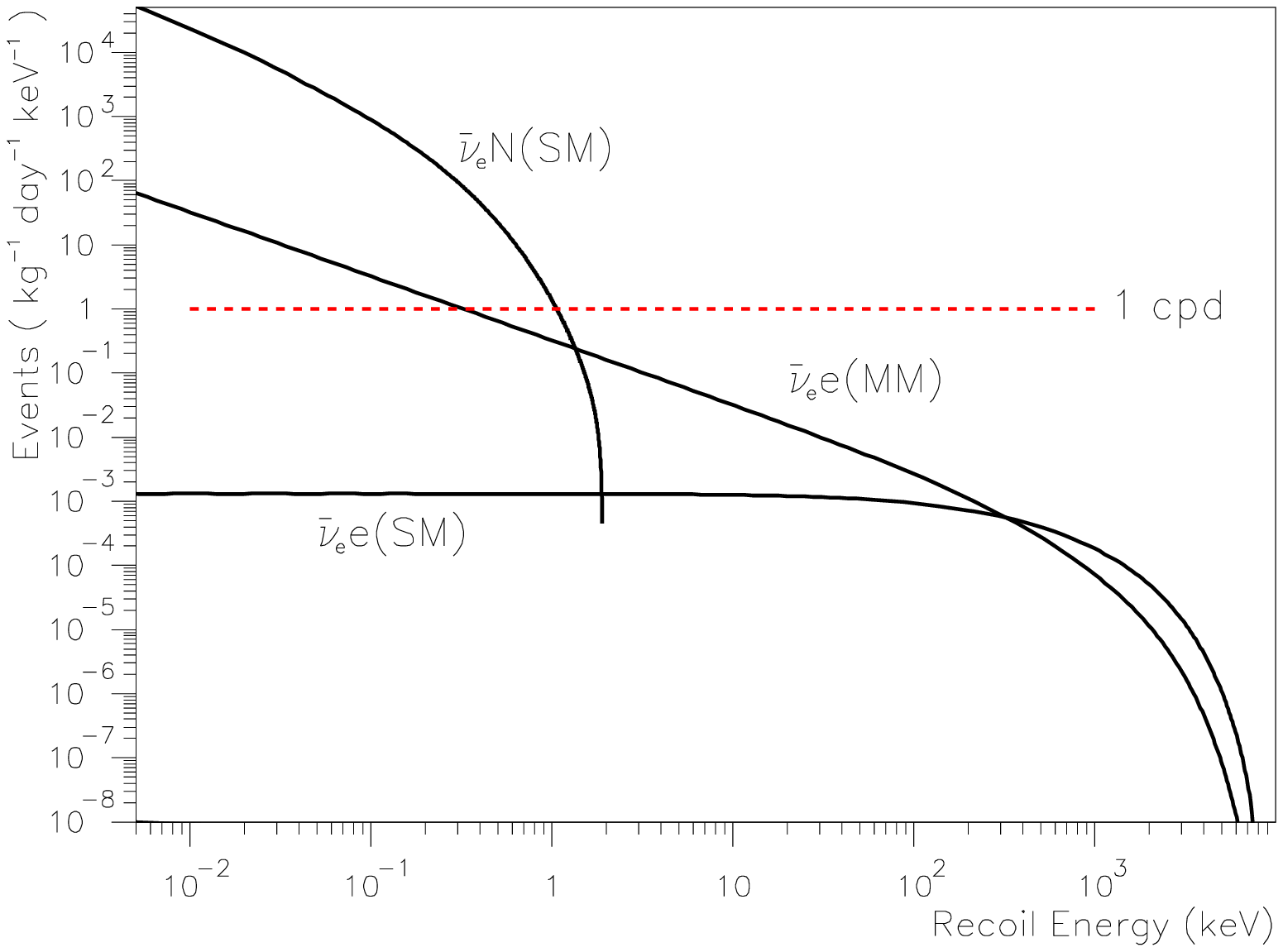}
\caption{\label{diffcs}
The differential cross-section of the various
neutrino interaction channels, at
KS-Lab with Ge as the target isotope. 
The background level of 1~cpd is also
shown.
}
\end{minipage}\hspace{2pc}%
\begin{minipage}{18pc}
\includegraphics[width=16pc]{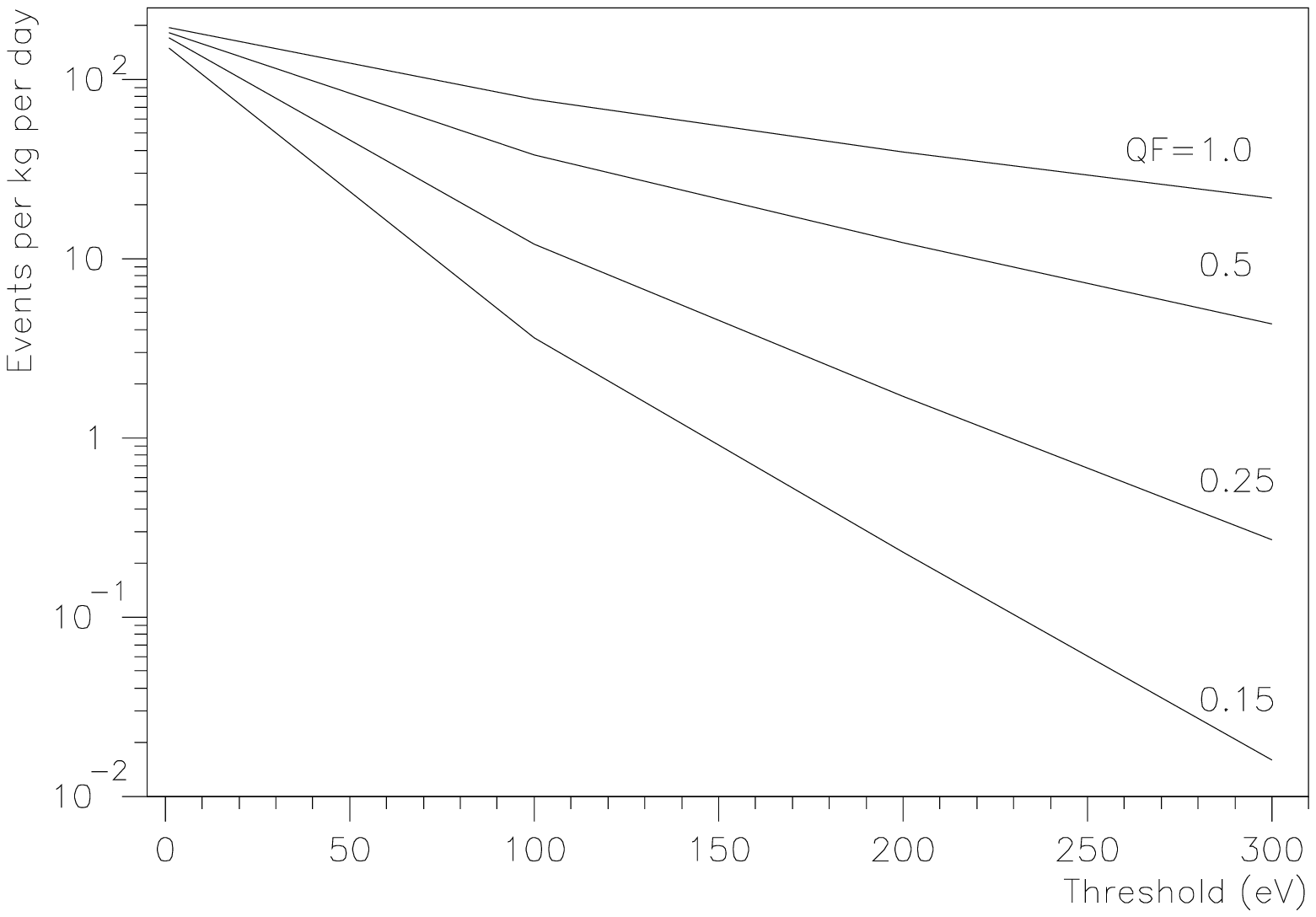}
\caption{\label{qf}
The variations of
the neutrino coherent scattering
event rates
versus threshold at different
quenching factors for
a 1~kg ULEGe detector at KS Lab.
}
\end{minipage}
\end{figure}

Nuclear power reactors provide 
powerful and controllable
source of electron anti-neutrinos,
and can serve as optimal tool for the studies
of neutrino-nucleus scatterings.
A research program on 
low energy neutrino physics\cite{ksprogram}
is intensely pursued by the TEXONO
Collaboration at the Kuo-Sheng (KS)
Nuclear Power Station in Taiwan.
The expected observable spectra due to
$\nuebar$-e and $\nuebar$-N scatterings
with Standard Model (SM) and magnetic
moment (MM) interactions at KS
are displayed in Figure~\ref{diffcs}.
The maximum nuclear recoil energy $\rm{T_{max}}$
in $\nuebar$-N coherent scatterings
is given by:
\begin{equation}
\rm{
T_{max} ~ = ~ \frac{2 E_{\nu}^2}{M_N + 2 E_{\nu}}
}
\end{equation}
which corresponds to
$\rm{T_{max} = 1.9~keV  }$
in the case of Ge target (A=72.6)
exposed to the typical reactor neutrino spectra.

High-Purity Germanium(HPGe) detectors have been widely and successfully
used in various areas of low energy neutrino physics
and cold dark matter searches.
These detectors are kilogram-scale in mass, and with
a detection threshold (``noise edge'') of
several keV.
A sensitive direct search of 
neutrino magnetic moments\cite{mmreview} was recently
performed 
with a 1~kg HPGe detector
at KS\cite{numagmom}. A physics threshold
of 12~keV and a background level of 
$\rm{\sim 1~ day^{-1} kg^{-1} keV^{-1} (cpd)}$
comparable to underground dark matter experiments
were achieved.
For ionization detectors like germanium,
the measure-able energy 
of nuclear recoil events
is only a fraction of their
energy deposited
due to charge recombination or 
``quenching'' at large dE/dx.
The expected event rates 
for neutrino-nucleus coherent scattering
at different threshold
and quenching factors (QF) at KS 
are depicted in Figure~\ref{qf}.
The QF for Ge
is typically 0.25 
in the several keV region, such that
the maximum measure-able energy
for nuclear recoil events 
is only about 500~eV.  

\begin{figure}[hbt]
\begin{minipage}{18pc}
\includegraphics[width=18pc]{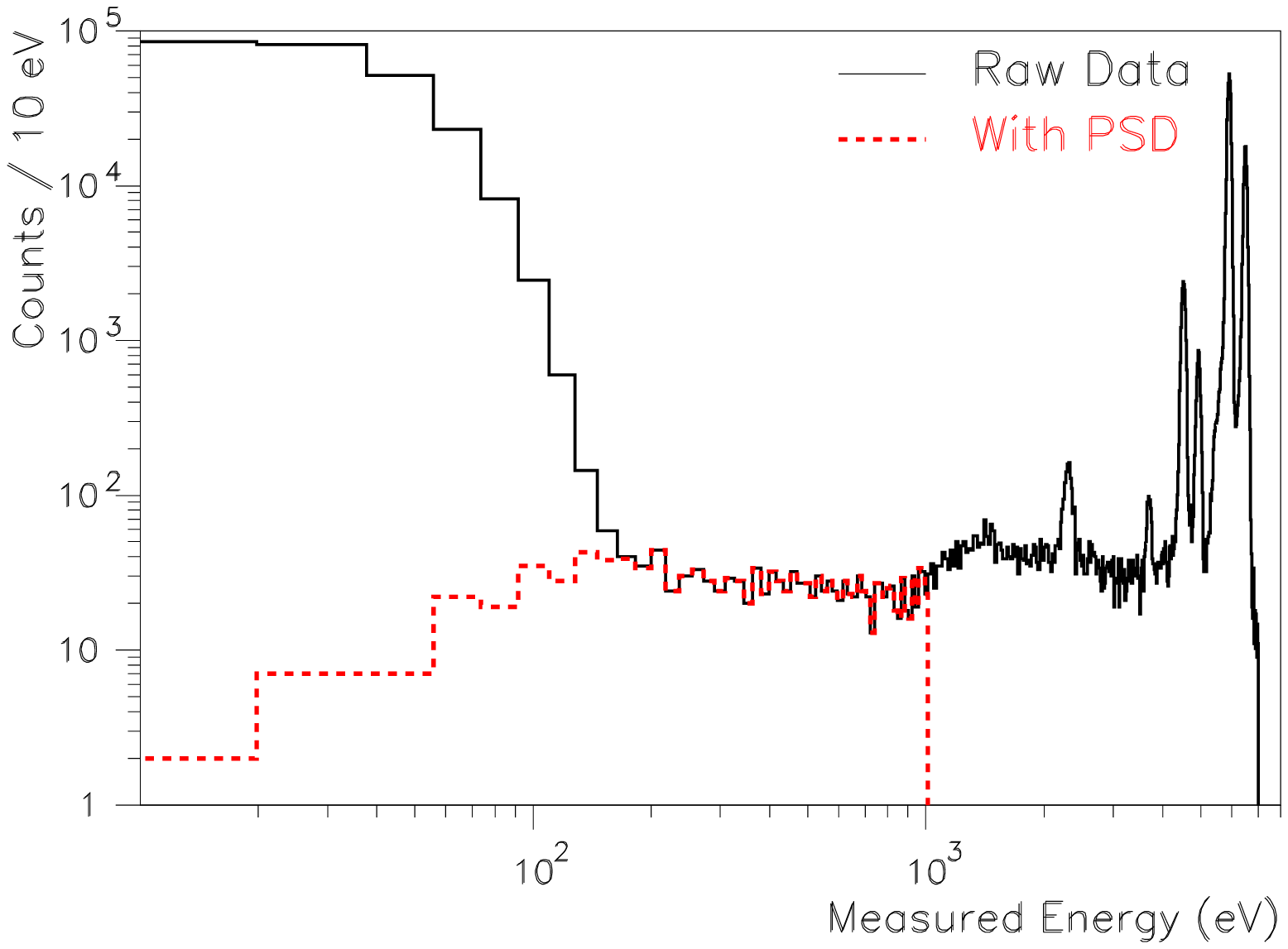}
\caption{\label{fe55}
Measured energy spectra with $^{55}$Fe
source with X-rays from Ti by the ULEGe
prototype.
A threshold of 100~eV was achieved,
and the electronic noise edge was suppressed
by PSD.
}
\end{minipage}\hspace{2pc}%
\begin{minipage}{18pc}
\includegraphics[width=18pc]{Figure4.eps}
\caption{\label{psd}
Pulse shape discrimination:
correlations of signals with
different electronic amplifications
and shaping times lead to
suppression of the noise edge.
}
\end{minipage}
\end{figure}

``Ultra-Low-Energy'' Germanium (ULEGe)
detectors, developed originally for soft X-rays
detection, are candidate technologies to meet
these challenges of probing into a previously
unexplored energy domain\cite{ulege}.
These detectors typically have
modular mass of 5-10~grams while detector array
of up to N=30 elements have been successfully
built. Various prototypes based on this
detector technology have been constructed.
As illustrations, the measured energy spectrum
with a 10~g ULEGe prototype
is depicted in Figure~\ref{fe55}.
Pulse shape discrimination (PSD)
criteria were applied
as illustrated in Figure~\ref{psd}
by correlating two output with
different electronics amplifications and 
shaping times.
Calibration was achieved by
external $^{55}$Fe X-ray sources (5.90 and  6.49~keV)
together with X-rays from
titanium (4.51 and 4.93~keV), calcium (4.01~keV),
sulphur (2.46~keV) and aluminium (1.55~keV). 
A random trigger uncorrelated to the detector
provided the zero-energy condition.
The electronic noise edge
can be  suppressed by PSD
and a threshold of 100~eV
was achieved.

\begin{figure}[hbt]
\begin{minipage}{18pc}
\includegraphics[width=16pc,height=12pc]{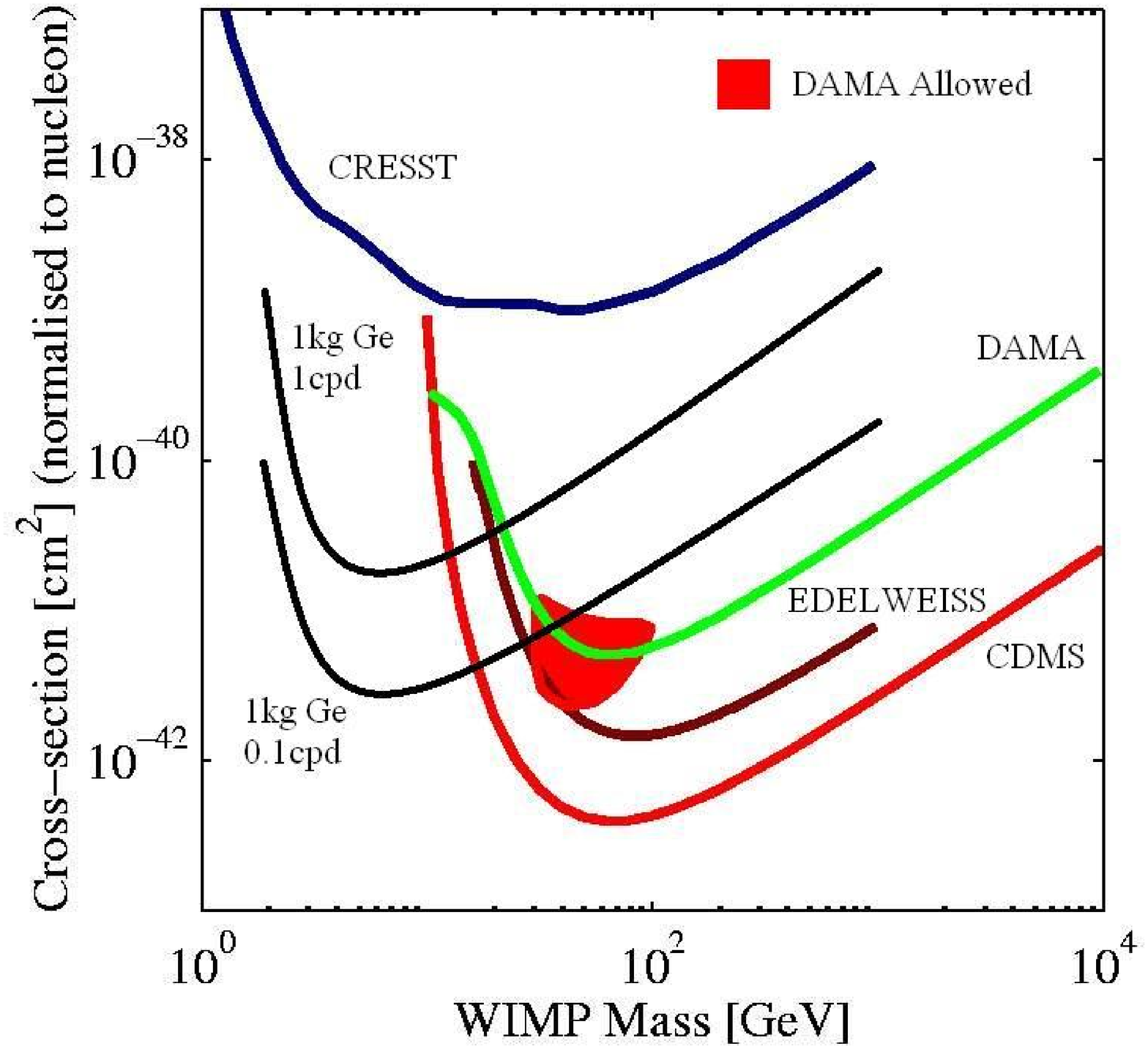}
\caption{\label{dmexplot}
Expected sensitivity region for
Cold Dark Matter searches using
a ULEGe detector with a total
mass of 1~kg.
}
\end{minipage}\hspace{2pc}%
\begin{minipage}{18pc}
\includegraphics[width=18pc,height=12pc]{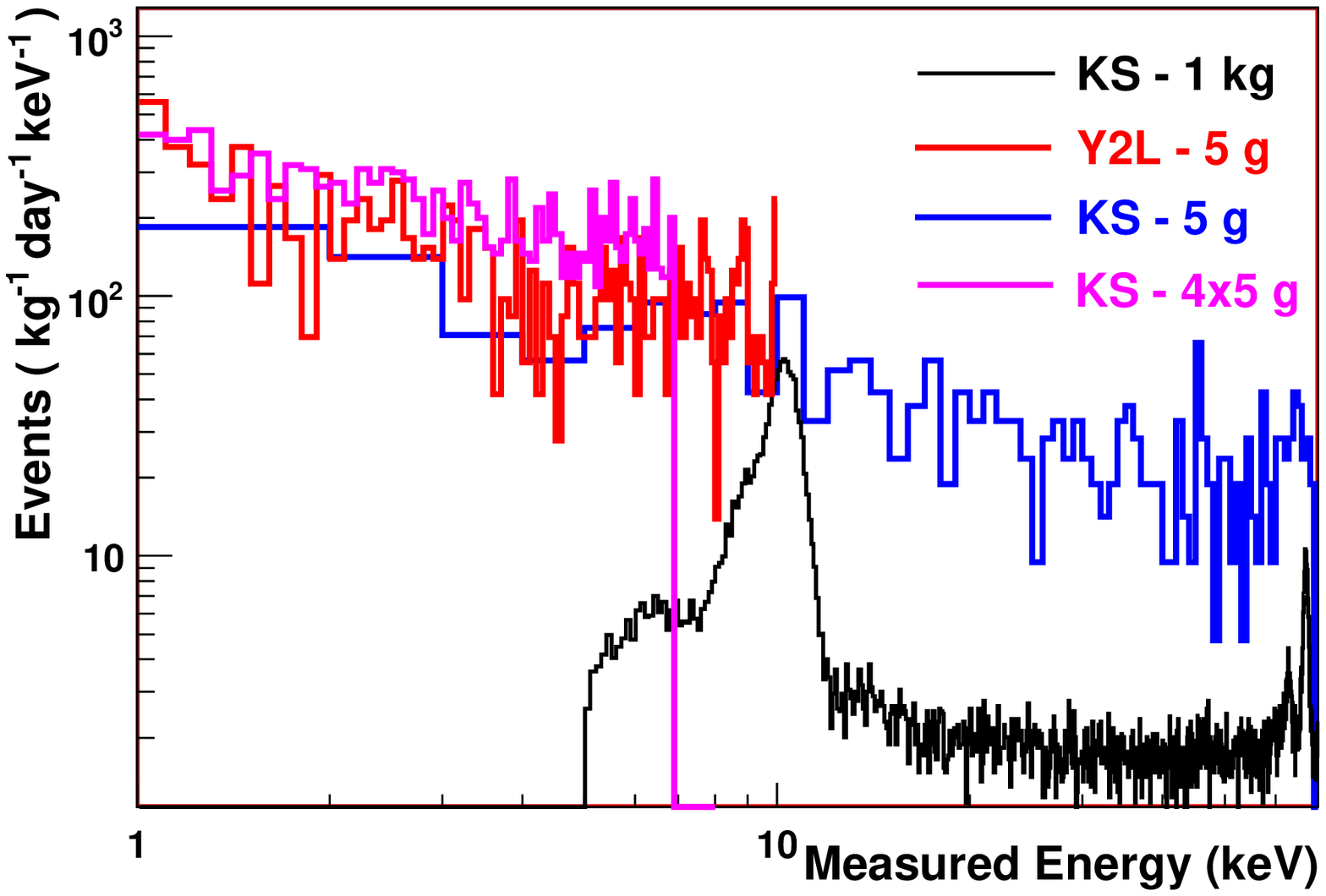}
\caption{\label{bkgspect}
Measured background spectra, normalized by mass,
at KS and Y2L with different Ge detectors. 
}
\end{minipage}
\end{figure}

The goal of the $\nu$-N coherent scattering
experiment is to develop a 
ULEGe detector
with a total mass of $\sim$1~kg and 
a modular threshold as low as 100~eV, with
a background level 
below 1~keV in the range of 1~cpd.
From Figure~\ref{qf} and
at the typical QF=0.25,
the event rate for such configurations at KS
will be 11 kg$^{-1}$day$^{-1}$, implying a signal-to-background
ratio of $>$22. A by-product of such an detector
would be to further enhance 
the searches of neutrino magnetic moment at
reactors. An improved sensitivity range down
to $\sim 10^{-11} ~ \mu_B$ can be expected\cite{mmreview}.
Such detector 
can also be used for Cold Dark Matter searches\cite{cdm}, 
probing the unexplored low WIMP-mass region, as indicated in
Figure~\ref{dmexplot}.

\begin{figure}[hbt]
\begin{minipage}{18pc}
\includegraphics[width=18pc,height=12pc]{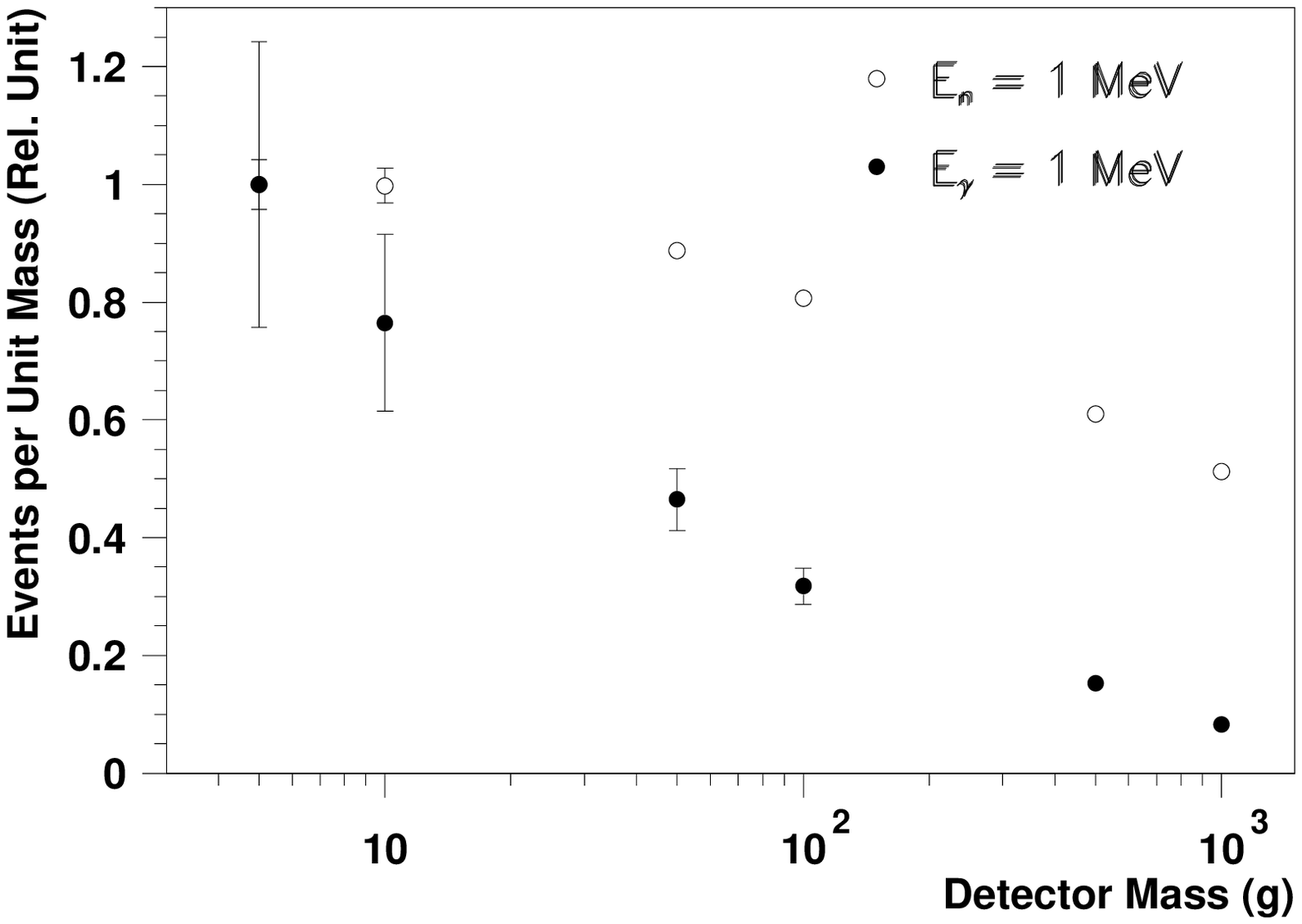}
\caption{\label{bkgmc1}
Simulated results on the variations
of background per unit mass versus detector
mass at the same external $\gamma$ and neutron
background level. 
}
\end{minipage}\hspace{2pc}%
\begin{minipage}{18pc}
\includegraphics[width=18pc,height=12pc]{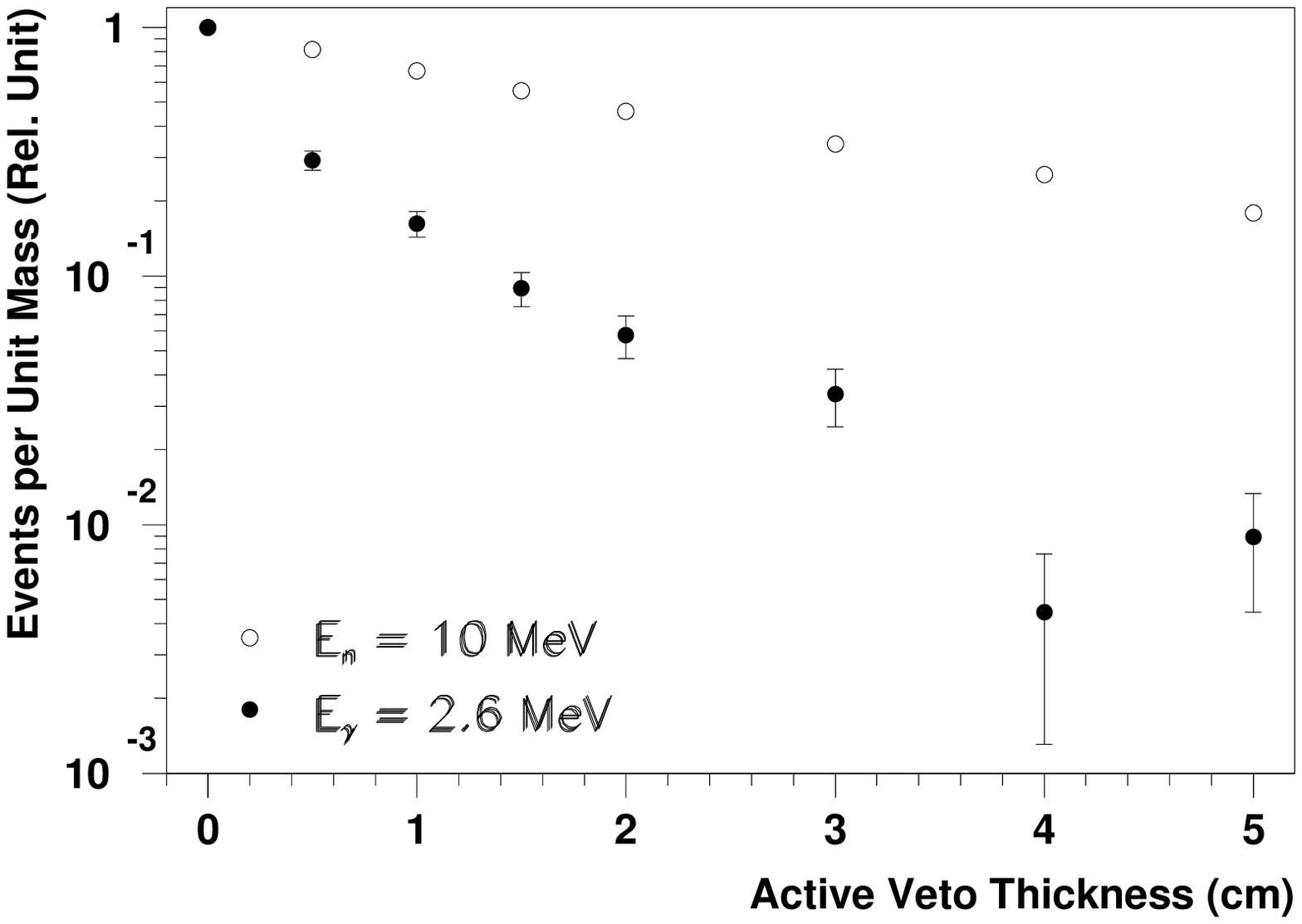}
\caption{\label{bkgmc2}
Simulated results on the background suppression
for $\gamma$ and neutron sources
at the inner ULEGe target due to an enclosing
structures of active Ge-vetos.
}
\end{minipage}
\end{figure}

An R\&D program towards realizations
of these experiments is being pursued.
A quenching factor measurement
for nuclear recoils in Ge with
sub-keV ionization energy 
is being prepared at a neutron beam facility.
Background studies are performed at both KS as
well as the Yang-Yang Underground Laboratory (Y2L)
in South Korea with the different prototypes.
The measured spectra 
above 1~keV 
on the various configurations are
presented in Figure~\ref{bkgspect},
while sub-keV background are under
intense studies.
The factor-of-ten difference in the 
mass-normalized background levels
between the 1~kg and 5~g detectors is
due to self-shielding of the detector.
Such effects are reproduced by 
realistic simulation studies
on the variations of $\gamma$ and 
neutron background levels
with detector mass, as
depicted in Figure~\ref{bkgmc1}.

Consequently, 
in the scaling-up to the kilogram mass-range,
the individual elements of the ULEGe 
should be assembled in a compact array 
in order to maintain the same
background level of $\sim$1~cpd achieved in
the 1~kg HPGe detector.
The ``segmented'' Ge technology with the
integrated circuitry approach is an
attractive alternative.
Configurations can be envisaged to
have active veto HPGe layers 
enclosing hermetically the ULEGe inner target.
Simulation results of Figure~\ref{bkgmc2} show
that a Ge-veto thickness of 3~cm can 
further suppress the background at the inner
target by about a factor of 30 and 3 for 
typical $\gamma$ and neutron background, respectively.
A prototype segmented ULEGe detector with a veto ring
and dual-readout channels
from both the signal and high-voltage
electrodes is being constructed.

\end{document}